\documentclass{emulateapj}
\usepackage{natbib}
\usepackage{amsmath}
\usepackage{graphicx}
\def \ni {\noindent}
\def\lsim{\mathrel{\rlap{\lower4pt\hbox{\hskip1pt$\sim$}}
    \raise1pt\hbox{$<$}}}                
\def\gsim{\mathrel{\rlap{\lower4pt\hbox{\hskip1pt$\sim$}}
    \raise1pt\hbox{$>$}}}                
\def\fml{$M/L_I$}
\def\fmls{$M/L_I$ }
\def\mbh{$M_{\bullet}$}
\def\mbhs{$M_{\bullet}$ }
\def\sersics{S\'{e}rsic }
\def\kms{$\text{km s}^{-1}$}

\def\apjl{ApJL}
\def\aj{AJ}
\def\mnras{MNRAS}
\def\apj{ApJ}
\def\araa{ARA\&A}
\def\nat{Nature}
\def\aap{A\&A}
\def\apjs{ApJS}

\defcitealias{kor88}{K88}
\defcitealias{kor96}{K96}
\defcitealias{gul09}{G09}



\begin{document}
\title{Orbit-Based Dynamical Models of the Sombrero Galaxy (NGC~4594)}
\shorttitle{Orbit-Based Dynamical Models of the Sombrero Galaxy}
\slugcomment{{\sc Accepted to ApJ:} 30 June 2011} 
\author{John R. Jardel\altaffilmark{1},
Karl Gebhardt\altaffilmark{1},
Juntai Shen\altaffilmark{2},
David Fisher\altaffilmark{3},
John Kormendy\altaffilmark{1},
Jeffry Kinzler\altaffilmark{1},
Tod R. Lauer\altaffilmark{4}, 
Douglas Richstone\altaffilmark{5}, and
K. G\"{u}ltekin\altaffilmark{5}
}

\altaffiltext{1}{Department of Astronomy, University of Texas at
Austin, 1 University Station C1400, Austin, TX 78712;
jardel@astro.as.utexas.edu, gebhardt@astro.as.utexas.edu
kormendy@astro.as.utexas.edu, kinzler@astro.as.utexas.edu}
\altaffiltext{2}{Key Laboratory for Research in Galaxies and Cosmology,
Shanghai Astronomical Observatory, Chinese Academy of Sciences, 80 Nandan 
Road, Shanghai 200030, China; jshen@shao.ac.cn}
\altaffiltext{3}{Laboratory for Millimeter-Wave Astronomy, 
Department of Astronomy, CSS 1204, University of Maryland, College Park, 
MD 20742-2421; dbfisher@astro.umd.edu}

\altaffiltext{4}{National Optical Astronomy Observatories, P. O. Box
26732, Tucson, AZ 85726; lauer@noao.edu}

\altaffiltext{5}{Dept. of Astronomy, Dennison Bldg., Univ. of
Michigan, Ann Arbor 48109; dor@astro.lsa.umich.edu, kayhan@umich.edu}

\begin{abstract}

We present axisymmetric, orbit-based models to study the 
central black hole, stellar mass-to-light ratio, and dark matter halo of
NGC~4594 (M104, the Sombrero Galaxy).  For stellar kinematics, we use
published high-resolution kinematics of the central region taken with the
\emph{Hubble Space Telescope}, 
newly obtained Gemini long-slit spectra of the major axis, 
and integral field kinematics from the SAURON instrument.
At large radii, we use globular cluster kinematics to trace the mass profile 
and apply extra leverage to recovering the dark matter halo parameters.  We
find a black hole of mass \mbh$=(6.6 \pm 0.4) \times 10^8 \, M_{\odot}$,
and determine the stellar $M/L_I=3.4 \pm 0.05$ (uncertainties are the 68\%
confidence band marginalized over the other parameters).
Our best fit dark matter halo is a 
cored logarithmic model with asymptotic circular speed 
$V_c=376 \pm 12 \text{ km s }^{-1}$ and core radius $r_c= 4.7 \pm 0.6$ kpc.
The fraction of dark to total mass contained within the half-light radius
is $0.52$.  Taking the bulge and disk components into account in our
calculation of  $\sigma_e$ puts NGC~4594 squarely on the $M$-$\sigma$ relation.
We also determine that NGC~4594 lies directly on the $M$-$L$ relation.

\end{abstract}

\keywords{galaxies: individual (M104, NGC~4594)---galaxies: kinematics and 
dynamics---galaxies: photometry }


\section{Introduction}

Most galaxies are thought to host supermassive black holes (SMBHs) at their
centers.  The masses of these SMBHs have been observed to correlate with 
several properties of their host elliptical galaxies and of the classical bulge
components of their host disk galaxies.  For example, \mbhs correlates with 
galaxy/bulge mass \citep{dre89,mag98,lao01,mcl02,mar03,har04}, luminosity
(the $M$-$L$ relation) \citep{kor93,kor95,kor01, gul09}, velocity dispersion
(the $M$-$\sigma$ relation) \citep{fer00,geb00,tre02, gul09}, and globular cluster
content \citep{bur10,har11}.  These and other, similar correlations 
suggest that galaxy 
formation and black hole growth are fundamentally linked.  To better understand
this interplay, accurate black hole masses are needed.

One challenge that limits the accuracy is 
the determination of the host galaxy's inclination.  
Projection effects are difficult to model and cause loss of
information, leading to systematic uncertainties.
Therefore, SMBHs in galaxies whose 
inclination is confidently known have the best chance of
being accurately and robustly measured.  Another issue that limits the accuracy
is the effect a dark matter halo has on the determination of SMBH mass.
\citet{geb09} show that orbit-based models can underestimate black hole mass
when dark matter is not considered in the modeling, however \citet{sch11} find
the effect is small when the black hole's sphere of influence is well-resolved.

NGC~4594 (M104, or the Sombrero Galaxy) is a nearly edge-on Sa type spiral 
with a 
prominent stellar disk and large, classical bulge \citep{kor04c}.  
The shape of this disk indicates that it (and thus
the entire galaxy) is inclined at an angle very close to $90^{\circ}$.
Throughout this paper we assume a distance to NGC~4594 of 9.8 Mpc,
calculated from surface brightness fluctuations \citep{ton01}.  Unless 
otherwise 
stated, all distance-dependent quantities are scaled to this value. 
\citet{ton01} use a value of $H_0=74 \text{ km s}^{-1}~\text{Mpc}^{-1}$ in
their distance determinations, however we compare
our \mbhs and $L_V$ to \citet{gul09} who adopt $H_0=70$ in their work.  
We therefore scale the \citet{gul09} distances down by 
$6\%$.  Black hole mass scales as \mbh$\propto D$ and luminosity as 
$L_V \propto D^{-2}$; these quantities are adjusted accordingly.

NGC~4594 was one of the first galaxies in which a black hole was detected, 
and it has a long history of study.
\citeauthor{kor88}(1988, hereafter
\citetalias{kor88}) first found evidence for a massive black hole of
\mbh$=5.4^{+11.8}_{-3.7} \times 10^8$ $M_{\odot}$ using only
ground-based observations.  With isotropic Jeans models, \citet{ems94b} measured
a black hole of mass \mbh$=(5.4 \pm 0.5) \times 10^8$ $M_{\odot}$.
Later, \citeauthor{kor96} (1996, hereafter 
\citetalias{kor96}) used high-resolution kinematics from the Faint Object
Spectrograph (FOS) on the \emph{Hubble Space Telescope} 
(\emph{HST})---the same data
set we include---to measure $\text{log } M_{\bullet}=8.8 \pm 0.5 \, M_{\odot}$.
This corresponds to a mass of $5.8^{+12.4}_{-4.0} \times 10^8 \, M_{\odot}$.
With isotropic models, \citet{mag98} obtained a value of
\mbh$=6.9^{+0.2}_{-0.1} \times 10^8$ $M_{\odot}$.
These values for \mbhs all lie towards the high mass end of the $M$-$\sigma$ 
and $M$-$L$ relations.  Massive SMBH 
measurements are frequently being revised, and we expect the confidently known 
inclination of NGC~4594 to lead to one of the 
more secure measurements of a high mass SMBH.

We present new Gemini spectroscopy of the major axis, as well as SAURON
integral field kinematics covering the central region of the galaxy.  
We also use high-resolution \emph{HST}/FOS kinematics of the nucleus 
and kinematics
derived from globular clusters at large radii.  We combine these kinematic
datasets with  \emph{HST} and ground-based photometry to run axisymmetric 
orbit-based models.
These models allow us to measure the black hole mass, stellar 
mass-to-light ratio, and dark matter halo of NGC~4594.  In addition, we
also recover information about the internal orbit structure of the galaxy.

\section{Data Reduction and Analysis}

Dynamical modeling requires as input the three-dimensional luminosity density 
distribution 
$\nu(r)$, as well as the line-of-sight velocity distribution (LOSVD) at 
many locations in the galaxy.  We use \emph{HST} and ground-based images for the 
photometry.  Our kinematics include high-resolution \emph{HST}/FOS spectra,
long-slit spectra from GNIRS on Gemini, SAURON integral field kinematics, 
and individual velocities of globular clusters (GCs).  We discuss 
each in turn.

\subsection{Photometry}

In order to cover a large enough dynamical range to have leverage on both the
central SMBH and dark halo, we use surface brightness profiles from \emph{HST}
and ground-based images.
The stellar disk of NGC~4594 dominates at intermediate radii on the major axis
causing the 
isophotes in this region to be substantially flattened.  This abrupt change in 
ellipticity introduces an additional challenge to the deprojection.  
Our standard technique is to assume that the surfaces of constant luminosity 
density $\nu$ are coaxial, similar spheroids \citep{geb96}.
Clearly the presence of a disk invalidates this assumption,
so we decompose the surface brightness into bulge and disk components,
deprojecting each separately so that our assumption holds for each component.
Afterwards, we re-combine the deprojected profile of each component
$\nu_{\mathrm{bulge}} + \nu_{\mathrm{disk}}$ and input the total $\nu(r)$ into our modeling program.

The bulge-disk decomposition fits directly to 
a projected image.  We construct a model disk by considering a \citet{ser68}
profile: 

\begin{equation}
\mu(R) = \mu_0 \mathrm{ exp}[-(R/R_0)^{1/n}]
\label{diskeq}
\end{equation}

\ni
where $\mu_0$ is the surface brightness at $R=R_0$ and $n$ is the \sersics
index.  For $n=1$, the profile is an exponential.  For inclinations other than 
$90^{\circ}$, the projection of our disk model is an ellipse.  By
specifying the inclination of the disk $i$, the axial ratio $b/a$ of the 
ellipse is given by $b/a=cos$ $i$ for a thin disk.

We construct many disk models by varying $\mu_0$, $R_0$, $i$, and $n$
(keeping $n$ close to 1). Each model is then subtracted from the image until
the residual brightness distribution has elliptical isophotes.
The remaining
light is assigned to the bulge component.  A 1D major axis bulge profile is
produced by averaging the bulge light in elliptical, annular isophotes.
Hence, we are left with an analytic disk model and a non-parametric bulge
model.  We identify the best bulge and disk models as those that 
minimize the rms residuals of the model-subtracted image.

\begin{deluxetable}{rrrrr}
\tablecaption{Summary of Disk Parameters Fit}
\tablewidth{0pt}
\tablehead{
  \colhead{Disk} & \colhead{$\mu_0$ (mag arcsec$^{-2}$)} & \colhead{$R_0$ (arcsec)} 
  & \colhead{$n$} & \colhead{$i$}}
  \startdata
   Outer 1    &   18.8  &  66.8  &  1.0  & 80 \\
   Outer 2 & 16.7 & 40.1 & 1.0 & 80 \\ 
   Nuclear & 20.4 & 4.1 & 1.1 & 83 \\
\enddata
\label{disktab}
\end{deluxetable}

\vskip 10pt

In addition to the obvious main disk, NGC~4594 hosts a well-studied 
nuclear disk (\citealt{bur86},\citetalias{kor88},\citetalias{kor96}) at small 
radii.  We fit the nuclear disk in
the \emph{HST} image and the main disk in the ground-based image.  
Because we fit directly to the images, dust lanes and object masking 
become important.  We keep a bad pixel list which instructs our code to ignore 
trouble spots.  Dust lanes are selected by eye, while SExtractor
\citep{ber96} is used to identify foreground stars, background galaxies, 
and globular clusters.

\subsubsection{HST Image}

To probe the nuclear region, we use a PSF-deconvolved \emph{HST} Wide Field 
Planetary Camera 2 (WFPC2) image (GO-5512; PI: Faber).  This image is
 presented in \citetalias{kor96} and provides 
an excellent view of the central region of the galaxy.
Centered on the PC1 camera, the image is
taken in the F547M filter, and has a scale of 0\farcs0455
pixel$^{-1}$ of the central $34 \arcsec \times 34 \arcsec$ of the galaxy.
The PSF deconvolution uses the Lucy-Richardson algorithm 
\citep{ric72,luc74} for 40 iterations
and is well-tested on WFPC2 images \citep{lau98}.
The best fit parameters from the bulge-disk decomposition are listed in 
Table \ref{disktab}.

NGC~4594 is also thought to have weak LINER emission \citep{ben06} and there is
a point source in the \emph{HST} image.  Furthermore, heavy dust absorption also makes
the determination of the central bulge surface brightness profile difficult 
for $R \lsim 0\farcs17$.  To deal with these issues, we extrapolate the bulge
surface brightness
$\mu_{\mathrm{bulge}}(R)$  inwards to $R = 0\farcs02$ with a constant slope 
fit to the region near $R=0\farcs17$.
Figure \ref{allcomps} shows the result of this extrapolation, as well as the
other components fit in the ground-based image.

\subsubsection{Ground-Based Image}
\label{ground-based image}

We obtained a wide-field, $I$-band image from the Prime Focus Camera on the 
McDonald 0.8 m telescope. This instrument provides a large unvignetted field 
of view (45$\times$45 arcmin$^2$ ) and a single CCD detector. Therefore, we 
can more robustly carry out sky-subtraction and accurately constrain the 
faint isophotes. The image is corrected for bias, flat field, and illumination
using standard routines in IRAF.

\begin{figure}[t]
\includegraphics[width=9cm,angle=0]{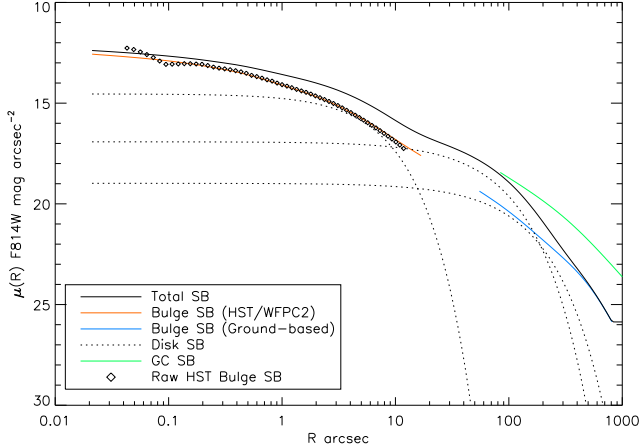}
\figcaption[allcomps_v2.ps]{Radial plot of all resulting components from
our bulge-disk decomposition along the major axis.  Dotted lines are the disk 
profiles with
parameters from Table \ref{disktab}.  The solid black line is the total surface
brightness (bulge + disks).  Solid colored lines are the bulge profiles, the 
red line is the result from fits to the \emph{HST} image, and the blue line is from 
the ground-based image. Diamonds indicate the raw \emph{HST} bulge profile before we
apply our dust correction and point-source removal.
The gap between the \emph{HST} and ground-based bulge 
profiles is interpolated over before deprojection.  Plotted in green is the 
globular cluster surface brightness profile, arbitrarily scaled to match the
stellar surface brightness at its innermost point.
\label{allcomps}}
\end{figure}

In our fit to the ground-based image, we ignore the central $20\arcsec$ due to
over-exposure and 
contamination from the nuclear disk.  We attempt fits in the region
$20\arcsec-900\arcsec$ with only one stellar disk given by Equation 
(\ref{diskeq}),  however these produce unacceptable residuals. 
Instead of modifying Equation (\ref{diskeq}), we add a second disk (in addition to the nuclear disk fit only in the \emph{HST} image).  
This approach is similar to 
the Multi-Gaussian Expansion technique used to model the light distribution
of bulges and ellipticals \citep{ems94a}. A summary of all components fit is 
given in Table \ref{disktab} and plotted in Figure \ref{allcomps}.

\vskip 10pt

\subsubsection{Globular Cluster Profile}

Globular clusters (GCs) are essentially bright test particles that allow us
to probe the potential at radii where the stellar light is 
faint.  They have been used in orbit-based models of other galaxies 
\citep{geb09,she10,mur11}.
To include them in our models, we use the GC number density profile
\citep{rho04} as an analog to the stellar density.  The number density profile
is converted to a surface brightness profile by arbitrarily adjusting the zero
point to match the stellar profile in log space. 

The green line in Figure \ref{allcomps} shows that the slope of the GC surface
brightness profile is different from that of the stars.  We run models using 
both the measured luminosity density distribution of the GCs and assuming that
of the stars.  We find significant preference for the measured GC profile.

\subsubsection{Bulge Profile}

Our bulge-disk decomposition returns a non-parametric form of the bulge
profile.  It is not necessary to have a parameterized bulge profile for
our dynamical models, however we fit a \sersics profile to 
ground-based bulge model using Equation (\ref{diskeq}).  The bulge is
well-fit by a \sersics function, with the rms residuals equal to
$0.08 \text{ mag arcsec}^{-2}$.  We measure 
$\mu_0=13.5 \text{ mag arcsec}^{-2}$,$R_0=0\farcs1$, and $n=3.7$.

We can convert the central surface brightness
$\mu_0$ and radius $R_0$ parameters to the more familiar ``effective'' 
parameters
$\mu_e$ and $R_e$.  The effective radius $R_e$ is given by $R_e=(b_n)^nR_0$
and the effective surface brightness $\mu_e=\mu_0+2.5\log(e) \,b_n$
\citep{mac03}.  The factor $b_n$ depends on $n$; an expansion for $b_n$ 
can be found in \citet{mac03}.  Applying these conversions, we obtain
$\mu_e=21.3 \text{ mag arcsec}^{-2}$ and $R_e=156.2\arcsec$.  

We obtain a simpler estimate for the half-light radius of the bulge from 
integration of the surface brightness profile; no fitting functions are 
required.  We estimate 
$R_e=117\arcsec \pm 12\arcsec$.  The integrated magnitudes are 
calculated for component $x$ by $L_x = 2 \pi \int I_x(r) r \, dr$.
This does not take into account the ellipticity of each component, so we
scale the luminosity by $L_x^{true} \approx (1 - \epsilon_x) L_x$ where 
$\epsilon_x$ is the ellipticity of each component, assumed to be constant with
radius.  The bulge profile is known to become rapidly circular for 
$r \gsim 100\arcsec$ \citep{bur86} so our procedure almost certainly underestimates
$M_{\mathrm{bulge}}$ and $B/T$.  These numbers are computed as a sanity check only, and
do not affect the dynamical models.

The absolute integrated magnitudes 
are $M_{\mathrm{disk}} = -21.4$ and $M_{\mathrm{bulge}} = -22.5$ in F814W, corrected for 
Galactic extinction along the line of sight \citep{sch98}.   Using the \emph{HST} 
calibration package SYNPHOT (described in detail below), we convert these
F814W magnitudes to $V$-band Vega magnitudes.  We obtain 
$M_{V,\mathrm{bulge}}=-22.1$ and $M_{V,\mathrm{disk}}=-21.0$.  
These structural parameters lie exactly on 
the fundamental plane for bulges and ellipticals as presented in 
\citet{kor09}.  Our integrated magnitudes
translate to a bulge-to-total ratio $B/T=0.73$ with the nuclear disk
contributing $1\%$ of the total light. This value of $B/T$ is lower than
previous measurements---\citet{kor11}
report $B/T=0.925 \pm 0.013$.  Our $B/T$, however, is in good agreement
with a recent measurement by \citet[model BD]{gad11}.  Regardless of the 
value of $B/T$, our dynamical models are unaffected, because we add all the 
bulge and disk light together again after the deprojection.

We do not explore the possibility of fitting an exponential stellar halo
in addition to a bulge and disk as \citet{gad11} do.  Our bulge-disk 
decomposition produces a non-parametric bulge profile which could in principle
be a combination of a \sersics bulge plus exponential halo.  
However, this resulting
profile is well-fit by a \sersics function with $n$ significantly larger than
1.  We therefore do not agree with the claim made by \citet{gad11} 
that the bulge of NGC~4594 is actually an exponential stellar halo.

\subsubsection{Deprojection}

We combine the \emph{HST} and ground-based bulge profiles by zero-pointing both to
F814W.  We calculate the F547M photometric zero point for the \emph{HST} image from
the SYNPHOT package in IRAF.  Spectral template fitting (Section \ref{stelkin})
shows that in the central region of the galaxy $ \gsim 85\%$ of the light 
comes from K6III stars.  We therefore convert the F547M zero point to F814W 
with SYNPHOT using the 
Bruzual Atlas\footnote{http://www.stsci.edu/hst/observatory/cdbs/bz77.html}
template for a K6III star.

Before deprojection, we extrapolate the one-dimensional profiles $\mu(R)$ 
with a constant slope to $R=1800\arcsec$.
The three disk profiles are then combined and deprojected via Abel inversion
in the manner described in \citet{geb96}.  We assume an inclination
of $i = 90^{\circ}$.  The inclinations of the combined 
disk components imply an ellipticity of $e=0.83$. 
Our composite bulge profile is 
deprojected in a similar fashion, assuming a constant ellipticity of 0.25
\citep{bur86}.  
We then add $\nu_{\mathrm{disk}}(r,\theta) + \nu_{\mathrm{bulge}}(r,\theta)$ 
to obtain the total luminosity density distribution $\nu(r,\theta)$ input to
our models.

\begin{figure}[t]
\includegraphics[width=9cm,angle=0]{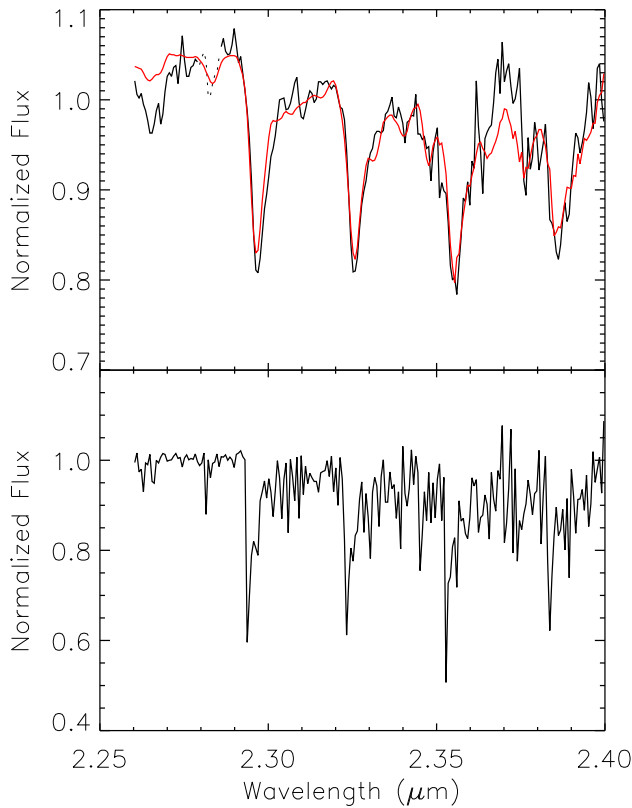}
\figcaption[spectrum]{Example Gemini spectrum.  Top: observed galaxy spectrum
(black) and best-fit LOSVD-convolved template star (red).  Dotted 
lines indicate regions of the spectrum ignored in the fit.  Bottom: 
spectrum of the template star.  The velocity dispersion of the LOSVD in this
fit is $\sigma=190 \pm 12 \,$\kms
\label{spectrum}}
\end{figure}
\vskip 10pt

The globular cluster luminosity density profile is obtained via a similar 
deprojection, but with the additional assumption of spherical symmetry.
The normalization of
the GC light profile is irrelevant, as our models fit only to the slope of 
the profile.

\subsection{Kinematics}

Kinematics for NGC~4594 come from four sources. The first uses near-IR
data from Gemini/GNIRS long-slit observations along the major
axis. These data were taken under good seeing conditions (around
0\farcs5) and has high $S/N$. The second set comes from the Faint Object
Spectrograph (FOS) on \emph{HST}
using the square aperture of $0\farcs21 \times0\farcs21$  and
is published in \citetalias{kor96}. The third set of data is from the 
SAURON instrument \citep{ems04}. The SAURON data for NGC~4594 have not
been published previously.  Individual velocities from globular clusters are
our fourth source of kinematics. These data are published in \citet{bri07}. 
We describe each dataset in detail.

\subsubsection{Gemini Kinematics}
\label{stelkin}

We use GNIRS \citep{eli06} on the Gemini South Telescope
to measure near-IR spectra of NGC~4594. The data were taken on 17 January,
2005. We placed the 150\arcsec$\times$0\farcs30 slit along the
major axis with the galaxy nucleus centered within the slit. We use a
spatial pixel size of 0\farcs15. With the 32l~l/mm grating in 3rd
order, we obtain a wavelength coverage of 19800--26200~\AA~at 6.4~\AA\
per pixel. Using sky lines, we measure a resolving power around 1700
or an instrumental dispersion of 75~km s$^{-1}$. The total on-target exposure
is 24 minutes, taken in 12, 2-minute individual exposures. Sky frames
of equal exposure are taken throughout.

From both setup images and images of telluric standards, we measure a
FWHM in the spatial direction of 0\farcs5, assuming a Gaussian
distribution. We use this PSF for the dynamical models.

We use a custom pipeline to reduce the GNIRS data; however, it
produces very similar results to the Gemini GNIRS reduction
package. The pipeline includes dark subtraction, wavelength
calibration for the individual exposures, sky subtraction,
registration and summing.

There is adequate signal to extract kinematics out to a radius of
45\arcsec. Figure \ref{spectrum} shows an example spectrum, where we plot 
the data
in black and the template convolved with the best-fit line-of-sight
velocity distribution (LOSVD). The velocity templates come from the
GNIRS spectral library \citep{win09}, where we select stars with a
range of types from G dwarf to late giant. The kinematic extraction
program performs a simultaneous fit to the LOSVD and relative weights
of the templates. This procedure is described in \citet{geb00}
and \citet{pin03}. We present these data in the form of Gauss-Hermite 
moments in Table \ref{gemtab}.

Figure \ref{moments} shows the kinematics derived from our analysis of the
Gemini spectra.  Between $1\arcsec$ and $5\arcsec$, $V$ rises and $\sigma$ 
drops.  This
is the result of the nuclear disk which becomes important at this radial 
range \citepalias{kor88}.  Beyond $10\arcsec$, we see similar behavior in 
$V$ and $\sigma$, it is caused by the main stellar disk.

\subsubsection{HST/FOS Kinematics}

\citetalias{kor96} present \emph{HST}/FOS kinematics of the nuclear region
of NGC~4594. The FOS has a
$0\farcs 21 \times 0\farcs21$ aperture. There are three pointings with
accurately known positions for NGC~4594 (GO-5512; PI: Faber).
The dynamical models include
the exact placement and aperture size for the FOS pointing \citepalias{kor96},
and use the \emph{HST} PSF \citep{geb00}.

\begin{deluxetable*}{rrrrrrrrr}
\tablecaption{Gemini Kinematics}
\tablewidth{0pt}
\tablehead{
	\colhead{$R$ arcsec} & \colhead{$V$ \kms} & \colhead{$\Delta V$ \kms} &
	\colhead{$\sigma$ \kms} & \colhead{$\Delta \sigma$ \kms} & 
	\colhead{$h_3$} & \colhead{$\Delta h_3$} & \colhead{$h_4$} & 
	\colhead{$\Delta h_4$}}
\startdata
0.00  &  19  &  14  &  253  &  16  &  -0.087  &  0.033  &  0.023  &  0.046  \\
0.15  &  -36  &  11  &  257  &  12  &  -0.008  &  0.042  &  -0.016  &  0.038  \\
0.30  &  -75  &  12  &  249  &  8  &  -0.053  &  0.048  &  -0.017  &  0.037  \\
0.52  &  -112  &  11  &  234  &  8  &  0.020  &  0.037  &  -0.047  &  0.035  \\
0.82  &  -144  &  12  &  221  &  9  &  0.060  &  0.038  &  -0.051  &  0.032  \\
1.20  &  -172  &  9  &  202  &  9  &  0.065  &  0.039  &  0.003  &  0.031  \\
1.73  &  -190  &  8  &  185  &  9  &  0.107  &  0.031  &  0.018  &  0.031  \\
2.40  &  -208  &  7  &  184  &  9  &  0.089  &  0.032  &  0.024  &  0.031  \\
3.30  &  -232  &  8  &  175  &  9  &  0.140  &  0.029  &  0.044  &  0.027  \\
4.57  &  -236  &  9  &  171  &  7  &  0.173  &  0.032  &  0.023  &  0.027  \\
6.45  &  -235  &  7  &  178  &  9  &  0.165  &  0.033  &  0.058  &  0.025  \\
8.77  &  -189  &  11  &  203  &  11  &  0.028  &  0.039  &  -0.007  &  0.036  \\
11.40  &  -171  &  10  &  192  &  10  &  0.071  &  0.033  &  -0.009  &  0.033  \\
14.32  &  -187  &  16  &  185  &  15  &  0.227  &  0.042  &  0.102  &  0.060  \\
17.70  &  -201  &  13  &  229  &  16  &  -0.023  &  0.042  &  0.020  &  0.050  \\
22.20  &  -228  &  12  &  198  &  14  &  0.125  &  0.040  &  0.052  &  0.047  \\
28.58  &  -235  &  11  &  194  &  14  &  0.122  &  0.042  &  0.036  &  0.051  \\
36.08  &  -285  &  6  &  141  &  10  &  0.046  &  0.041  &  0.032  &  0.034  \\
44.40  &  -277  &  7  &  149  &  9  &  0.094  &  0.042  &  0.101  &  0.039  \\
-0.15  &  45  &  12  &  240  &  15  &  -0.026  &  0.034  &  0.002  &  0.041  \\
-0.30  &  112  &  11  &  243  &  15  &  -0.052  &  0.036  &  -0.016  &  0.038  \\
-0.52  &  130  &  10  &  224  &  15  &  -0.080  &  0.043  &  0.014  &  0.037  \\
-0.82  &  162  &  8  &  212  &  10  &  -0.068  &  0.049  &  0.019  &  0.030  \\
-1.20  &  176  &  7  &  206  &  11  &  -0.066  &  0.047  &  0.065  &  0.031  \\
-1.73  &  205  &  7  &  190  &  12  &  -0.087  &  0.043  &  0.022  &  0.037  \\
-2.40  &  226  &  7  &  173  &  12  &  -0.101  &  0.037  &  0.053  &  0.039  \\
-3.30  &  244  &  9  &  187  &  11  &  -0.113  &  0.039  &  0.037  &  0.037  \\
-4.65  &  246  &  8  &  184  &  12  &  -0.130  &  0.035  &  0.025  &  0.040  \\
-6.15  &  237  &  9  &  220  &  14  &  -0.133  &  0.042  &  0.094  &  0.035  \\
-8.40  &  221  &  9  &  211  &  13  &  -0.191  &  0.051  &  0.079  &  0.036  \\
-10.95  &  162  &  9  &  209  &  14  &  -0.093  &  0.048  &  0.028  &  0.043  \\
-14.25  &  178  &  9  &  198  &  14  &  -0.092  &  0.059  &  0.046  &  0.042  \\
-19.20  &  204  &  10  &  198  &  17  &  -0.166  &  0.060  &  0.094  &  0.048  \\
-24.52  &  207  &  11  &  193  &  17  &  -0.045  &  0.057  &  0.029  &  0.052  \\
-29.33  &  241  &  11  &  182  &  16  &  -0.149  &  0.054  &  0.092  &  0.050  \\
-36.15  &  271  &  11  &  180  &  12  &  -0.084  &  0.046  &  0.053  &  0.042  \\
-45.15  &  274  &  10  &  141  &  9  &  -0.053  &  0.041  &  -0.029  &  0.028  \\

\enddata
\label{gemtab}
\tablecomments{Kinematics along the major axis of NGC~4594.  Gauss-Hermite 
moments were derived from the LOSVDs that are the input to the dynamical 
models.}

\end{deluxetable*}

\subsubsection{SAURON Kinematics}

We also include SAURON integral field kinematics The SAURON data are
from a single pointing exposing on the central region, taken in the low
resolution setting of the instrument \citep{bac01}.  In addition to
$V$ and $\sigma$, the SAURON data also include the higher order
Gauss-Hermite moments $h_3$ and $h_4$.  Details of the data reduction and
analysis can be found in \citet{bac01, ems04}.

Our modeling code fits to the entire LOSVD rather than its moments, so
we reconstruct LOSVDs from the Gauss-Hermite moments.  We create 100
Monte Carlo realizations of a non-parametric LOSVD from the
uncertainties in the Gauss-Hermite parameters of each SAURON bin
\citep{geb09}.  The 1433 reconstructed SAURON LOSVDs are spatially
sampled more finely than our modeling bins.  We therefore average the
SAURON data to match our binning by weighting according to the
uncertainties in the LOSVDs.

We re-construct Gauss-Hermite moments from the combined SAURON LOSVDs
for plotting purposes only.  Figure \ref{moments} shows these moments
near the major and minor axes.  The major axis $V$ for the SAURON data
is significantly lower than that measured for the Gemini data.  The
reason for this is that SAURON data are binned to match the gridding
of our model bins.  Near the major axis, the bins range in polar angle
from $\theta=0-11^{\circ}$.  These bins are described by a single LOSVD
constructed by averaging individual LOSVDs which sample the region at smaller
spatial scales.  Thus, the average LOSVD contains contributions from 
LOSVDs as much as $\theta=11^{\circ}$ above the major axis.

\subsubsection{Globular Cluster Kinematics}

At large radii, we use individual globular cluster velocities
published in \citet{bri07} to derive LOSVDs.  The data contain
positions and radial velocities for 108 globular clusters in NGC~4594.
We discard the innermost 14 GCs as there are too few GCs inside $R
\lsim 130\arcsec$ to reconstruct an LOSVD in the inner parts of the
galaxy. Assuming axisymmetry, the positions of the GCs are folded
about the minor and major axes.  In order to preserve rotation, we
flip the sign of the velocity for all GCs that are folded about the
minor axis.  The GCs are then divided into annular bins extending from
$\theta=0^{\circ}$ to $90^{\circ}$ at radii of
$131\arcsec$, $214\arcsec$, $350\arcsec$, $574\arcsec$, and
$941\arcsec$ with roughly 20 GCs per bin.  

Within each spatial bin, we calculate the LOSVD from the discrete GC
velocities by using an adaptive kernel density estimate adapted from
\citet{sil86} and explained in \citet{geb96}.  Each LOSVD contains 15
velocity bins.  The velocity bins are highly correlated for the GCs, and there
are likely only a few degrees of freedom per LOSVD.
The $1$-$\sigma$ uncertainties in the LOSVDs are
estimated through bootstrap resamplings of the data
\citep{geb96,geb09}.  

\begin{figure}
\includegraphics[width=9cm,angle=0]{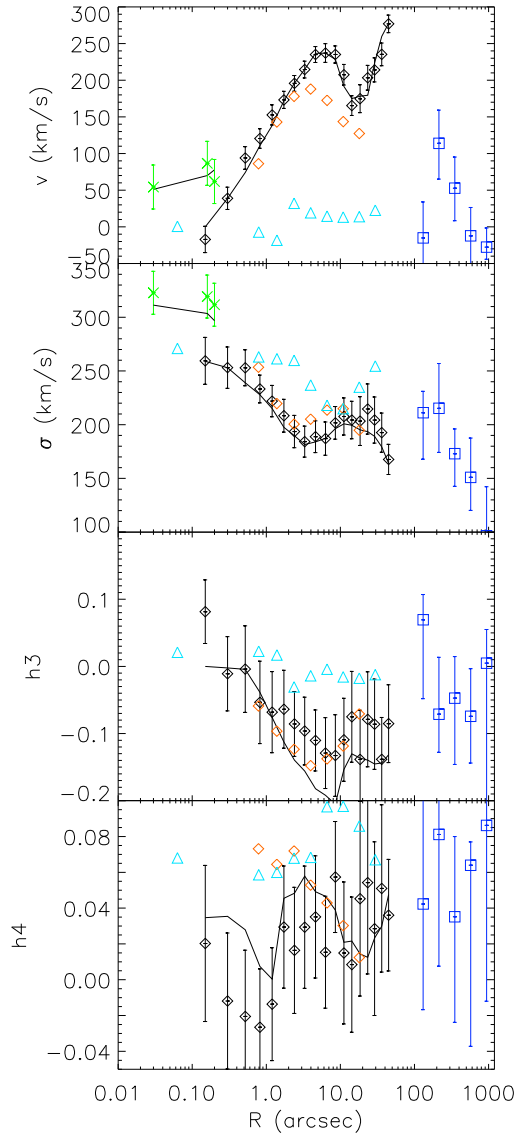}
\figcaption[moments]{Gauss-Hermite moments for NGC~4594 from various
sources.  
Black diamonds with error bars are from Gemini long-slit observations along
the major axis.
Red diamonds are from SAURON data near the major axis.  Light blue triangles
are SAURON data near the minor axis.  Green crosses are the 3 \emph{HST} data points,
and dark blue squares are from the globular clusters.  Solid lines are the 
result of our best fit model.
\label{moments}}
\end{figure}

We compute Gauss-Hermite moments from the GC LOSVDs---again for
plotting purposes only---and show these in Figure \ref{moments}. 
Uncertainties are calculated by fitting moments to each resampling of the LOSVD
during the bootstrap.
The GC kinematics resemble the minor axis stellar kinematics in many of
the panels.  For example, their dispersions appear to be an
extrapolation of the minor axis velocity dispersions.  There is slight
rotation (small $h_3$) and possible evidence of radial anisotropy
(positive $h_4$).

\vskip 10pt

\section{Dynamical Models}

The dynamical models rely on the orbit superposition technique first
developed by \citet{sch79}.  We assume axisymmetry and match
the luminosity density profile and kinematics of the galaxy to those
reconstructed from an orbit library.
The library is populated with orbits carefully chosen to sample 
E, $L_z$, and the third, non-classical integral $I_3$.

The code used in this paper is described in 
\citet{geb00,geb03}, \citet{tho04,tho05}, and \citet{sio09}.  
Similar axisymmetric codes are presented in \citet{rix97}, \citet{vdm98}, 
\citet{cre99}, and \citet{val04}. Van den Bosch et al. (\citeyear{vdb08})
present a 
fully triaxial Schwarzschild code. The basic outline of our code 
is as follows: ({\it i.}) convert the luminosity density distribution 
$\nu (r)$ into the stellar density $\rho (r)$ via an assumed stellar 
mass-to-light ratio \fml.  ({\it ii.}) Add to this density the
contribution from a black hole of mass \mbhs and a dark matter halo with 
density profile $\rho_{DM} (r)$.  ({\it iii.}) Calculate the potential $\Phi$
associated with this density distribution and integrate a large number of 
orbits (typically $\sim$ 20,000) over many dynamical times.  ({\it iv.}) 
Assign a weight $w_i$ to each orbit and determine the 
$w_i$ values by minimizing the $\chi^2$ difference between the observed 
kinematics and luminosity density of the galaxy and those resulting from the
PSF-convolved orbit library, subject also to the constraint of maximum entropy.

We maximize the entropy-like
quantity $\hat{S} \equiv S-\alpha \chi^2$ where $S$ is 
the Boltzmann entropy and $\alpha$ controls the relative weight of
$S$ or $\chi^2$. For small values of $\alpha$, 
reproducing the observed kinematics  becomes
unimportant, and the models act to only maximize entropy.
As $\alpha$ increases,
maximizing entropy becomes less important, and more weight is given to
matching the observations.  In practice,
we start with a small value of $\alpha$ and gradually increase it until 
$\chi^2$ asymptotes.  The interested reader may see \citet{sio09} or
\citet{she10} for more details.

Our model grid consists of 19 radial and 5 azimuthal bins covering a radial
range of 0\farcs03 to 1800\arcsec spaced logarithmically.  Additionally,
we use 15 velocity bins to describe our LOSVDs.
We incorporate the effects of seeing by convolving the light distribution
for each orbit with a model PSF before comparing with data \citep{geb00}. 
We approximate the PSF as Gaussian with a FWHM of either $0\farcs94$, 
$0\farcs5$ or $0\farcs09$ depending on whether the data are from SAURON,
Gemini, or \emph{HST} observations respectively.  The convolution extends to a radius
of $10 \times \text{FWHM}$.

We run over 8,500
models with different values of the model parameters \fml, \mbh, and 
$\rho_{DM}$.  We use $\Delta \chi^2$ statistics to determine the best fit 
parameter values and their uncertainties.
Models whose values of $\chi^2$ are within $\Delta \chi^2=1$ of the minimum
for a given model parameter (marginalized over the others) define 
the $1$-$\sigma$ or 68\% confidence band of that parameter.

\begin{figure*}[ht]
\centerline{\includegraphics[width=15cm,angle=0]{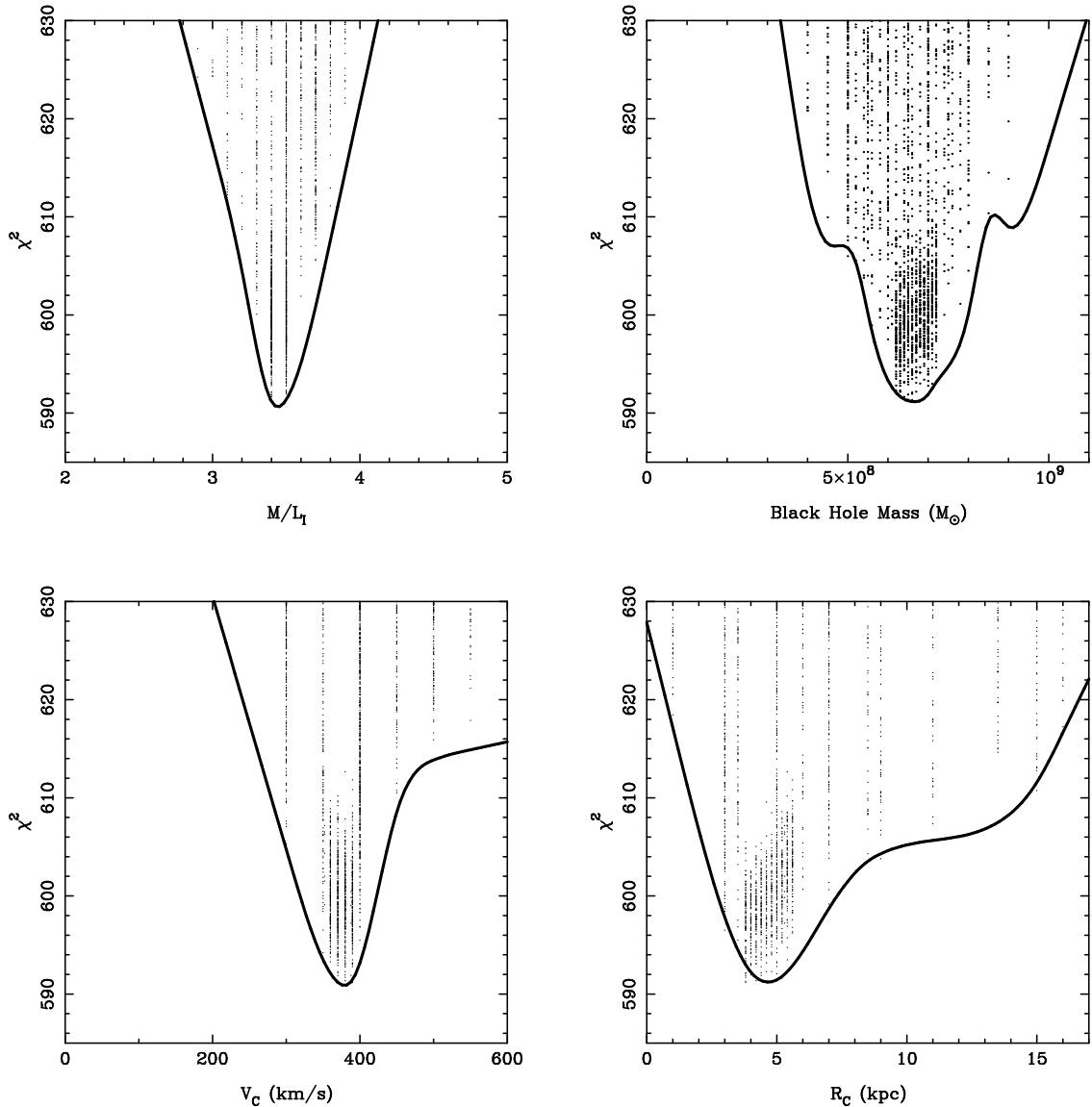}}
\figcaption[plotac2.ps]{$\chi^2$ as a function of the four modeled
parameters---\fml, \mbh, $V_c$, and $r_c$.  Every dot represents a single 
model.  The solid line is a smoothed fit to the minimum, which represents
the marginalized values.
\label{reschi2}}
\end{figure*}

\subsection{Model Assumptions}

Our fiducial density profile is a combination of stellar mass, dark matter, and
a central SMBH:

\begin{equation}
\rho(r,\theta) = \frac{M}{L} \nu(r,\theta) + \rho_{DM}(r,\theta) + M_{\bullet} \delta(r)
\label{denseq}
\end{equation}

\ni
where \fmls is the stellar mass-to-light ratio, assumed constant
with radius and $\delta(r)$ is the Dirac delta function.  The angle $\theta$ is
the angle above the major axis.
While $\rho_{DM}$ can in principle be a function of $\theta$, we do not 
consider flattened models.  We assume a spherically symmetric,
logarithmic halo of the form:

\begin{equation}
\rho_{DM}(r)=\frac{V_c^2}{4\pi G} \frac{3r_c^2+r^2}{(r_c^2+r^2)^2}
\label{loghaloeq}
\end{equation}

\ni
This profile is cored for radii $r \lsim r_c$ and produces a flat rotation 
curve with circular speed $V_c$ for $r \gg r_c$.
It has two free parameters, $r_c$ and $V_c$, which are varied
in the fitting process.  Including \fmls and \mbh, this brings the total 
number of model parameters to four.

Recently, \citet{geb09} have shown that the inclusion of a dark matter halo
can significantly affect modeled BH masses.  To test for this,
we run a smaller suite of models without a dark halo.

\section{Results}

Our best-fit values for the four model parameters are $M/L_I=3.4 \pm 0.05 \,
\frac{M_{\odot}}{L_{\odot}}$, 
\mbh$=(6.6 \pm 0.4) \times 10^8 \, M_{\odot}$,
$V_c=376 \pm 12 \text{ km s }^{-1}$, and 
$r_c= 4.7 \pm 0.6$ kpc.  Figure \ref{reschi2} shows the $\chi^2$ minima
around each of the model parameters.  Each dot represents a single model and
the solid curve is a smoothed fit to the minimum. The points of the solid 
curve at $\Delta\chi2 = 1$ above the minimum determine the $1$-$\sigma$ 
confidence limits on the parameters.
All four model parameters have well-behaved $\chi^2$ curves with sharp, 
well-defined minima.  This allows robust determination of the
model parameters with small $1$-$\sigma$ uncertainties.

\begin{figure*}[t]
\centerline{\includegraphics[angle=0,width=15cm]{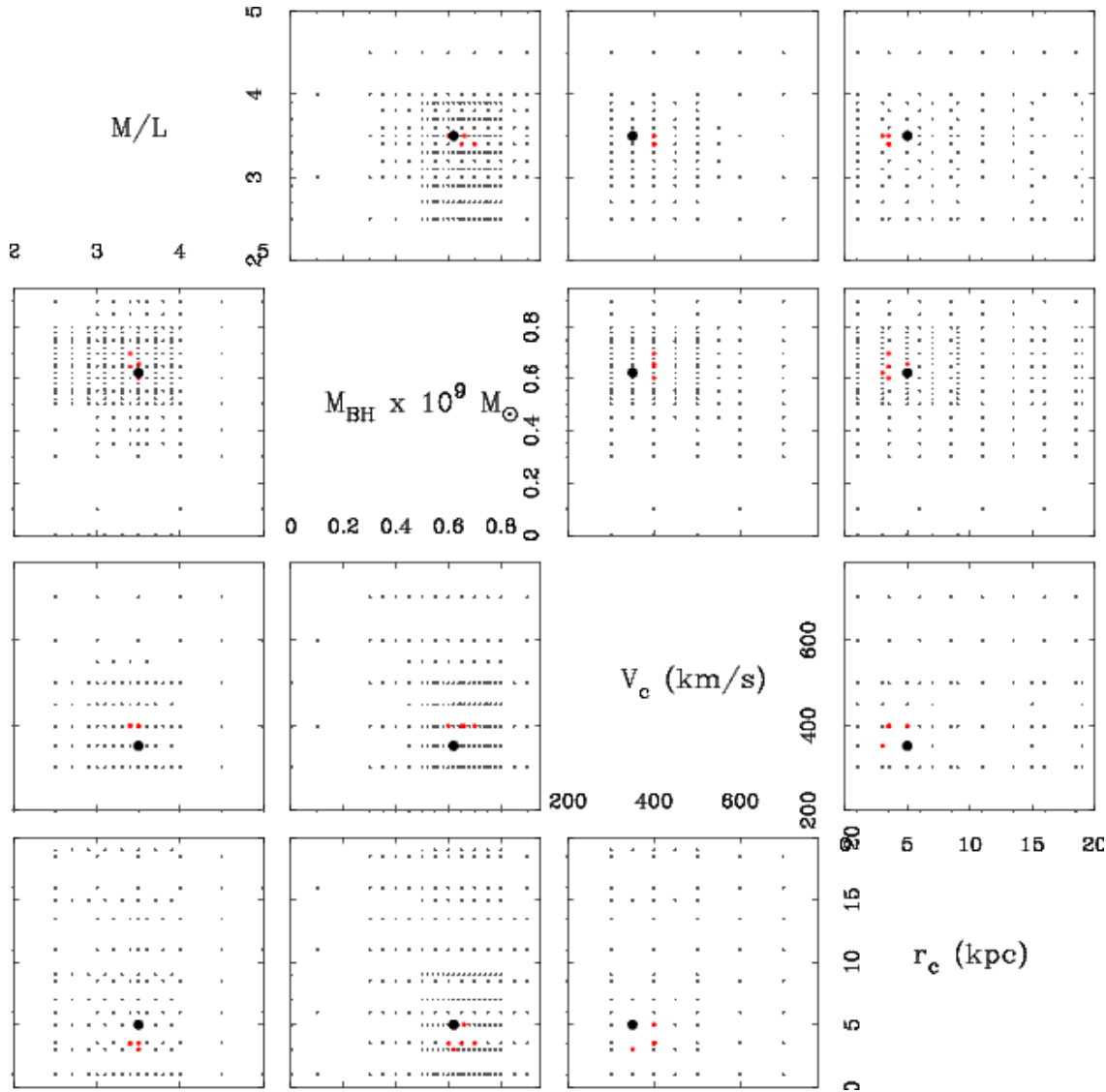}}
\figcaption[plotall4594.eps]{Correlation plots among the 4 parameters.
Each dot represents a single model.  Red dots are within the 95\% confidence
band and large black dots are within the 68\% confidence band for an individual
parameter
\label{corrplot}}
\end{figure*}

The uncertainties we present are derived strictly from $\Delta \chi^2$
statistics.  The $1$-$\sigma$ error bars on quoted parameters correspond to 
models within $\Delta \chi^2 =1$ of the minimum value.  
Systematic effects are likely to contribute in addition to this quoted 
uncertainty.  While in general for other
galaxies one of the biggest sources of 
systematic uncertainty is inclination, for NGC~4594 inclination
uncertainties are unimportant.
Other sources of uncertainty may include effects due to non-axisymmetries, but 
these are
likely small or zero since only the most massive ellipticals are thought to be
significantly triaxial \citep{bin78,kor82,tre96}.  For more on
systematic uncertainties, the reader is referred to \citet{geb03} and 
\citet{gul09b, gul09}.

Figure \ref{corrplot} shows correlations among the four model parameters.  
Plotted are the different projections of the four dimensional parameter 
space; every small dot corresponds to a model run.  Red dots are models that 
lie within $\Delta \chi^2=4$ of the minimum, and large black dots are within 
$\Delta \chi^2=1$.  There appears to be a slight correlation between \fmls
and \mbh---much less severe
than in M87 \citep{geb09}.  Not surprisingly, the high resolution of our 
\emph{HST} kinematics is able to break the degeneracy between \mbhs and \fml.
We discuss this further below.
The dark halo parameters do not show any obvious correlation, indicating the 
GC and stellar kinematics were able to break the degeneracy 
usually observed between these two parameters.

Our best fit model has (unreduced) $\chi^2=582.6$. It is non-trivial
to calculate the number of degrees of freedom $\nu_{DOF}$.  Roughly, 
$\nu_{DOF}=N_{LOSVD} \times N_{bin}$, however there are complicated 
correlations
between velocity bins \citep{geb03}.  With this crude estimate for 
$\nu_{DOF}$, our best fit model has reduced $\chi^2_{\nu}=0.6$.

We compare the modeled value of our stellar mass-to-light ratio with
that obtained from evolutionary population synthesis models 
\citep{mar98,mar05}.  We adopt values of 10 Gyr and 0.1 for the stellar 
age and metallicity of NGC~4594 \citep{san06} and use these to derive the
predicted I-band \fmls from the Maraston models.  For a Salpeter IMF with
stellar masses drawn from the range $0.1-100 M_{\odot}$, this analysis yields
\fml$ = 3.99$.  If instead the stars obey a Kroupa IMF drawn from the same 
range, then \fml $=2.58$.  We multiply these by a factor of 1.096 corresponding
to $A_I=0.099$ to correct for Galactic extinction along the line of sight
\citep{sch98} to obtain \fml $=2.83$ and \fml $=4.37$.  Our 
dynamically-derived stellar \fml $=3.4 \pm 0.05$ falls nicely between these
two values.

\begin{figure}[t]
\includegraphics[width=9cm]{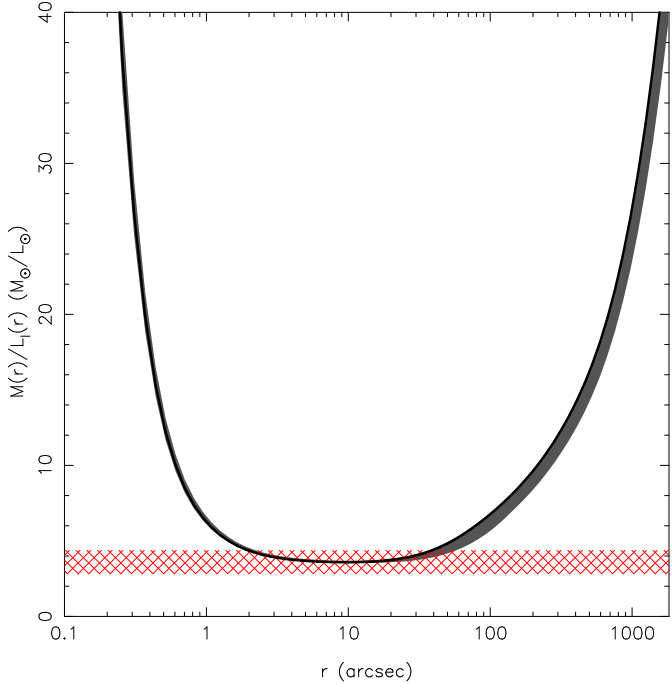}
\figcaption[n4594ml.ps]{Local dynamical mass-to-light ratio for the best
fit model.  The gray band indicates the 68\% confidence band, as 
determined from the limits placed on the 4 model parameters.
The red cross-hatched region indicates the extinction corrected 
stellar \fmls derived from population synthesis models.
\label{ml}}
\end{figure}

In Figure \ref{ml}, we plot the total mass-to-light ratio as a function
of radius for our best fit model with $1$-$\sigma$ uncertainties (gray 
region).  The red cross-hatched region represents the range in stellar \fmls 
from stellar population models described above.  Total \fmls rises near the 
center of the galaxy due to the 
contribution of the supermassive black hole.  As we go out in radius, the 
stars become more important to the total mass over roughly the range 
$5\arcsec$ to $50\arcsec$.  Here the total \fmls approaches both our 
dynamically determined \fmls and the range derived from stellar population
models.  Past $50\arcsec $ \fmls once again rises due to 
the importance of the dark halo.

\begin{figure}[t]
\includegraphics[width=9cm]{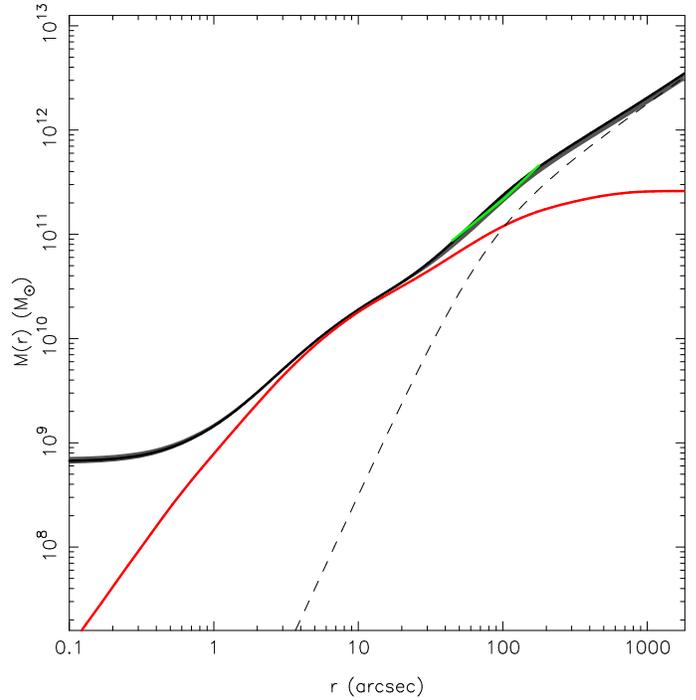}
\figcaption[mr.ps]{Mass enclosed within spherical shells for our best fit model
and 68\% confidence region.
The red line is the 
stellar mass profile while the black line and surrounding confidence region
represent the total mass (black hole + stars + DM).  The dashed line
is our best-fit dark matter halo.  Green indicates the 
mass profile derived in \citet{kor89} from gas rotation.
\label{mr}}
\end{figure}

Figure \ref{mr} plots the 
enclosed mass of each component as well as the total mass of the galaxy.
At our innermost bin, the total mass is almost two 
orders of magnitude greater than the stellar mass, meaning we are probing the
black hole's sphere of influence quite well.
The green line plotted is the mass profile \citet{kor89} derive from
their gas rotation curve.  It agrees well with the total mass distribution
derived here.

\vskip 10pt

\subsection{Models without Dark Matter}

We run 189 models with no dark halo.  In these models, we exclude the globular
cluster data and use only the stellar kinematics. Correspondingly, the number
of degrees of freedom impacting the unreduced $\chi^2$ are proportionately
fewer.   We measure a black hole
mass of \mbh$=(6.6 \pm 0.3 ) \times 10^8 \, M_{\odot}$ and stellar mass-to-light ratio of 
$M/L_I=3.7 \pm 0.05$. The minimum unreduced $\chi^2=628$, proving models 
without a dark halo are a worse fit.  

We do not see the dramatic change that 
\citet{geb09} see in M87 where the inclusion of a DM halo causes their 
determination of \mbhs to
double.  Instead, our results mirror those of \citet{she10} in NGC 4649 
where the inclusion of a DM halo does not significantly change the 
modeled \mbh.
The likely explanation for this behavior is the inclusion of high resolution 
\emph{HST} kinematics in both NGC 4649 and NGC~4594. \citet{sch11} find the same 
effect for a larger sample of galaxies.  Whenever the data have high enough
resolution to resolve the black hole's radius of influence 
$R_{\mathrm{inf}} \sim G M_{\bullet}/\sigma^2$, dark matter has no significant
effect on the determination of \mbh.  For NGC~4594 we measure 
$R_{\mathrm{inf}} \simeq 57 \text{ pc} \simeq 1\farcs2$.  We use 
\emph{HST}/FOS kinematics
whose central pointing has a PSF of $0\farcs09 \simeq 0.08 \, R_{\mathrm{inf}}$.
Additionally, the light profile
of NGC~4594 is more centrally concentrated than that of M87. These factors 
combine to allow a more accurate 
determination of \mbh, removing the freedom that the models have to trade
mass between \mbhs and \fml.  This is evidenced by the lack of correlation 
among \mbhs and \fmls in Figure \ref{corrplot}.

\subsection{Orbit Structure}
\label{orbitstructure}

Having already determined the orbital weights that provide the best fit
to the data, we reconstruct the 
internal unprojected moments of the distribution function. We perform this
analysis on our best fit model and the models that define the 68\% confidence
region (over all combinations of the four model parameters), yielding
internal moments at each grid cell.

\begin{figure}[t]
\includegraphics[width=9cm,angle=0]{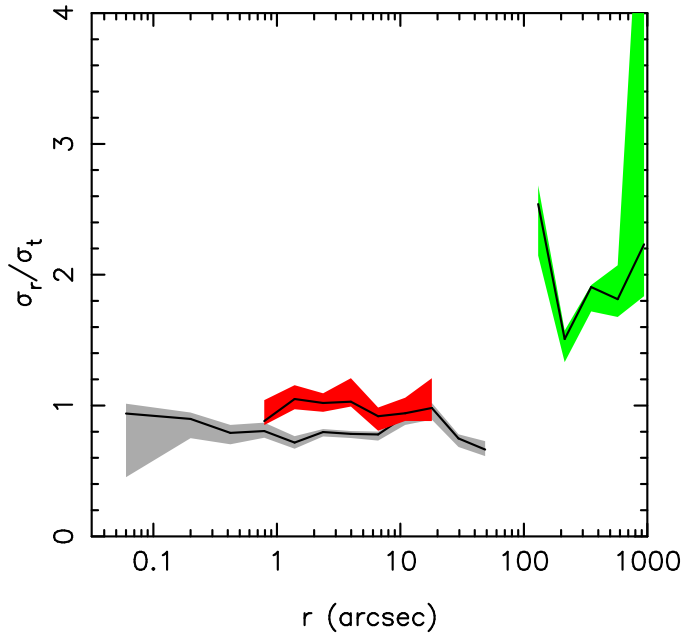}
\figcaption[vr_vt_use.ps]{Radial run of the ratio of the radial to tangential 
components of the velocity dispersion tensor.  Shaded ares represent 
68\% confidence regions, with gray indicating stars near the major
axis, red meaning stars near the minor axis, and green representing GCs 
averaged over all angles.
\label{vr_vt}}
\end{figure}

We define the tangential velocity dispersion to be
$\sigma_t \equiv \sqrt{\frac{1}{2}(\sigma_{\phi}^2+\sigma_{\theta}^2)}$
where $\sigma_{\phi}$ is actually the second moment, containing
contributions from both streaming and
random motion in the $\phi$ direction.  Figure \ref{vr_vt} shows the radial
 run of the ratio
$\sigma_r/\sigma_t$.  The second moment of the DF is tangentially biased
where the disk is important (gray region) as expected but mostly is isotropic 
elsewhere.  The red region plots $\sigma_r/\sigma_t$ for stars near the minor
axis, showing almost perfect isotropy.  
The green region indicates that at large radii, globular cluster 
kinematics show significant radial anisotropy. We discuss the implications of 
this in Section \ref{gcsection} below.

\section{Discussion}

\subsection{Black Hole-Bulge Correlations}

We discuss the position of NGC~4594 on the $M$-$\sigma$ and 
$M$-$L$ relations, as defined by \citeauthor{gul09} (2009b, hereafter
\citetalias{gul09}) and compare our values of the
correlation parameters to previous measurements.  We calculate the 
effective velocity dispersion $\sigma_e$ similar to \citetalias{gul09}

\begin{equation}
\sigma_e^2 \equiv \frac{{\int}^{R_e}_{R_{\mathrm{inf}}} (V^2(R) + \sigma^2(R)) I(R) dR}{{\int}^{R_e}_{R_{\mathrm{inf}}} I(R) dR}
\label{sigmae}
\end{equation}

\ni
where $V(R)$ is the rotational velocity and $I(R)$ is the surface brightness
profile.
This makes $\sigma_e$ essentially the surface-brightness-weighted second moment.
Instead of integrating from the center of the galaxy ($R=0$) as 
\citetalias{gul09} did, we integrate from $R > R_{\mathrm{inf}}$ to ensure that
we do not bias $\sigma_e$ 
with the high dispersion near the black hole.  Our outermost kinematic data 
point is at $R=45\arcsec$, thus we have a gap in kinematic coverage between
$45\arcsec < R < R_e=114\arcsec$.  The velocity dispersion near the end of our
long-slit data is dropping sharply, however $V$ may still contribute to the 
integral for
$R>45\arcsec$.  To investigate this, we use the gas rotation curve 
presented in \citet{kor89} which extends well beyond $R_e$.  Truncating the
integral at $R=45\arcsec$ gives $\sigma_e=292 \text{ km s}^{-1}$ while using the
extended rotation curve yields $\sigma_e=297 \text{ km s}^{-1}$.  

The problem with this definition of $\sigma_e$ is that it includes 
a contribution from the rotation of the disk.  It has been shown
that black hole mass does not correlate with disk properties 
\citep{kor11} so this is not ideal.  However, to compare with \citetalias{gul09} we
must be consistent in our calculation of $\sigma_e$.  We therefore quote
this value of $\sigma_e$ when we compare to the $M$-$\sigma$ relation 
determined by \citetalias{gul09}.  As we expect black hole mass to track 
bulge quantities, disk contribution to $\sigma_e$ is likely to add a source of
intrinsic scatter to spiral galaxies in the $M$-$\sigma$ relation.  In fact,
spiral galaxies are observed to have larger scatter about $M$-$\sigma$
than ellipticals of similar $\sigma_e$.

We also discuss some possible alternatives to $\sigma_e$ where we attempt
to remove the disk contribution.  One option is to remove $V^2(R)$ from 
Equation (\ref{sigmae}) altogether. Bulges are known to rotate, however,
\citep{kor82} and this will likely underestimate $\sigma_e$.
This crude calculation gives $\sigma_e=200 \text{ km s}^{-1}$.

Another option is to assume some degree of bulge rotation a priori.  If we
assume NGC~4594 rotates isotropically \citep{kor82}, then its flattening
determines its position on the $V/\sigma-\epsilon$ diagram \citep{bin78}.
\citet{kor82b} shows that the relation 

\begin{equation}
\frac{V}{\sigma} \approx \sqrt{\frac{\epsilon}{1-\epsilon}}
\label{v/sigma}
\end{equation}

\ni
approximates the isotropic rotator line to roughly $1\%$ accuracy.
We use our value of the bulge ellipticity $\epsilon=0.25$ in Equation 
(\ref{v/sigma}) and assume this value of $V/\sigma$ applies globally to 
the entire bulge.  We then use our measured dispersion profile $\sigma(R)$
to determine the bulge velocity $V_{\mathrm{bulge}}(R)$.  Using these quantities, we
determine $\sigma_e=230 \text{ km s}^{-1}$.  We compare this to the kinematics
listed in \citet{kor82}.  These data include long-slit spectra taken at a 
position angle parallel to the major axis, but $30\arcsec$, $40\arcsec$, and
$50\arcsec$ above it.  From these data, it is apparent that the bulge $\sigma$
off the major axis is roughly constant at $\sim 220 \text{ km s}^{-1}$.  The
rotation velocity rises from 0 to $100 \text{ km s}^{-1}$ at large radii. We
estimate the luminosity-weighted mean $V \sim 50 \text{ km s}^{-1}$.
Adding this in quadrature to the constant bulge $\sigma=220 \text{ km s}^{-1}$
gives $\sigma_e \approx 226 \text{ km s}^{-1}$.  This estimate does not 
contain any rotation
from the disk, and is consistent with our determination of $\sigma_e=230 
\text{ km s}^{-1}$ obtained by assuming a constant $V/\sigma$.

\begin{figure*}[t]
\centerline{\includegraphics[width=15cm,angle=0]{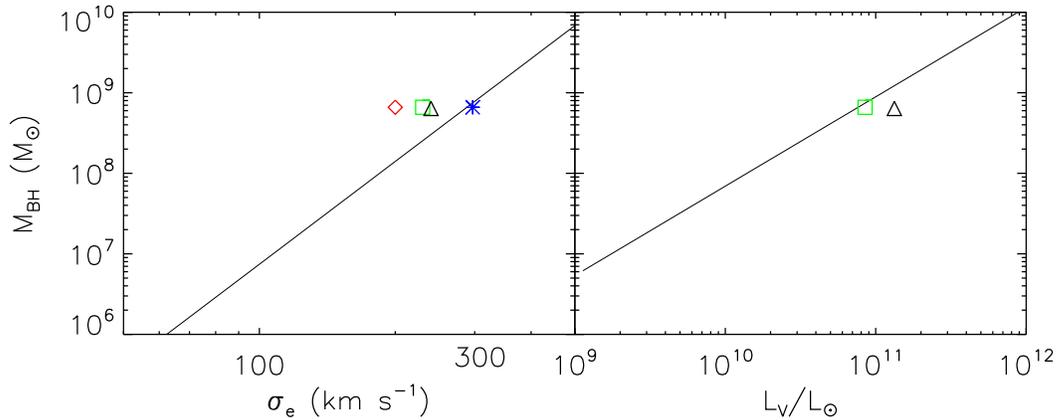}}
\figcaption[msigma.eps]{Position of NGC~4594 on the \citetalias{gul09}
$M$-$\sigma$ and $M$-$L$ relations.  The plot of $M$-$\sigma$ (left) shows
the three ways we calculate $\sigma_e$ as well as the value from 
\citetalias{gul09} (black triangle).  In order of increasing $\sigma_e$
we plot $\sigma_e$ with no rotation (red diamond), $\sigma_e$ assuming
a value of $V/\sigma$ (green square) and $\sigma_e$ as in 
\citetalias{gul09} (blue asterisk).  For the $M$-$L$ relation (right)
we plot the \citetalias{gul09} value (black triangle) along with our
measurement (green square).
\label{msigma}}
\end{figure*}

\begin{figure}[t]
\includegraphics[width=9cm,angle=0]{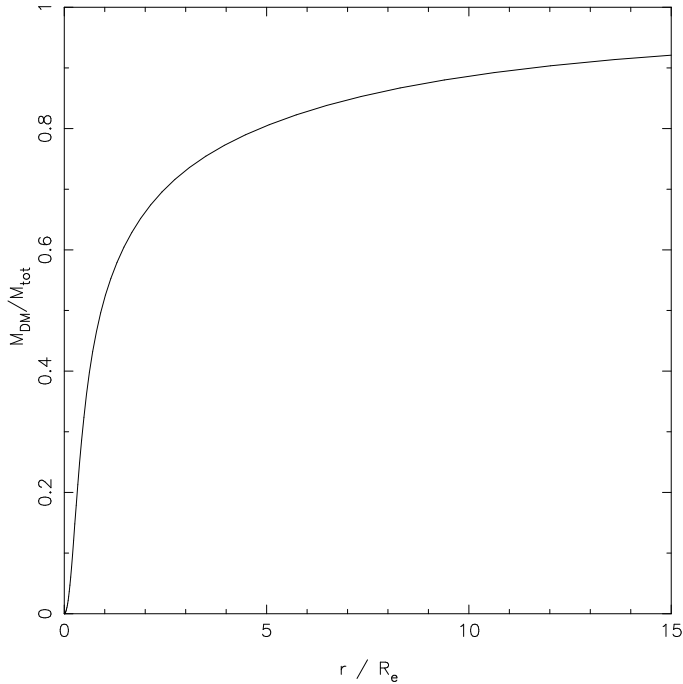}
\figcaption[fracdm.ps]{Fraction $M_{DM}/(M_{\star}+M_{DM})$ of enclosed
mass that is dark matter as a function of radius.
\label{fracdm}}
\end{figure}

Our black hole mass \mbh$=(6.6 \pm 0.4) \times 10^8 \, M_{\odot}$ agrees
nicely with that of \citetalias{gul09} which uses the \citetalias{kor88} value.
In fact, when corrected for their
various distance determinations, most values of \mbhs in the literature
agree quite well 
(\citetalias{kor88}, \citealt{ems94b}, \citetalias{kor96}, \citealt{mag98})
despite the many modeling techniques and datasets used.  This is likely 
due to the high degree of isotropy as evidenced in Figure \ref{vr_vt}, 
deduced from the $V/\sigma-\epsilon$ diagram \citep{kor82}, and noted
in \citetalias{kor88}.

Figure \ref{msigma} plots the position of NGC~4594 on the \citetalias{gul09}
$M$-$\sigma$ and $M$-$L$ relations.  We plot each determination of 
$\sigma_e$ in the left-hand panel.  Straightforward application of Equation
(\ref{v/sigma}) leads to a value of $\sigma_e$ that falls directly on the
\citetalias{gul09} $M$-$\sigma$ line (blue asterisk).  Next closest is the 
method of calculating $\sigma_e$ by assuming a value of $V/\sigma$ 
(green square).  This 
point lies $0.44$ dex above the \citetalias{gul09} line, however this is still
within the estimated scatter.  The calculation of $\sigma_e$ that ignored
all rotation is, not surprisingly, farthest from the \citetalias{gul09} 
line.  Calculation of the relevant quantities for comparison with the
$M$-$L$ relation is straightforward, and we plot our value of \mbhs and $L_V$
(green square) along with that from \citetalias{gul09} in the right-hand 
panel.

\subsection{Globular Clusters}
\label{gcsection}

As demonstrated in section \ref{orbitstructure}, we find significant radial 
anisotropy in the globular clusters.  It is interesting that the stellar 
kinematics at smaller radii do not show this feature.  This difference in 
orbital properties combined with the difference in their light profiles might
suggest the GCs and stars are two distinct populations of tracer particles.
This could also indicate the two populations have different formation 
scenarios.

Unfortunately, there is 
no radius in the galaxy where we have simultaneous coverage of both stellar and
GC kinematics.  Thus, we are unable to test whether the stellar orbits become 
more radial in the $\sim 50\arcsec$ between where the stellar kinematics run 
out and the GCs begin.  However, since the light profiles of both populations
are significantly different (Figure \ref{allcomps}) there is no 
reason to assume they should share similar orbit properties.

Figure \ref{vr_vt} shows the globular clusters in NGC~4594 are radially 
anisotropic ($\sigma_r/\sigma_t >1$) over roughly the radial range 
$100$-$1000\arcsec$ (approximately 1-10 $R_e$).
Previous studies of the GC systems of galaxies have found their velocity
ellipsoids to be isotropic \citep{cot01,cot03}.  However, these studies
used spherical Jeans modeling
instead of the more general axisymmetric Schwarzschild code we use.

\citet{rho04} determine with high confidence that the color 
distribution of the GC system in NGC~4594 is bimodal.  This may indicate 
different 
subpopulations of GCs with different orbital properties that formed at
different epochs in the galaxy's history. In our analysis, we make no 
distinction between red and blue subpopulations.  We use the light profile and
kinematics of all available GCs, regardless of color.  However, since we 
use different sources for our kinematics and photometry data, there is
the possibility that each source draws from a different GC subpopulation.

\subsection{Dark Halo}

The parameters $V_c$ and $r_c$ of our model dark halo imply a central
dark matter density of $\rho_c=0.35 \pm 0.1 \, M_{\odot} \text{ pc}^{-3}$
Using an improved Jeans modeling technique, \citet{tem06} model
NGC~4594 and find a dark matter halo with central density 
$\rho_c=0.033 \, M_{\odot} \text{ pc}^{-3}$, ten times lower than our value.  
They, however, measure a larger stellar \fml$_V=7.1 \pm 1.4$ in the bulge.

We plot the fraction of enclosed mass that is dark matter as a function of
half-light radius $R_e$ in Figure \ref{fracdm}.  At $1$ $R_e$ there is already
a roughly $50$-$50$ mix of stars and DM.  Inside of $R_e$ the dark 
matter still contributes a non-negligible fraction to the total mass 
content.  

In a study measuring dark matter properties in $1.7 \times 10^5$ local
($z<0.33$) early-type galaxies from the Sloan Digital Sky Survey, 
\citet{gri10} find a correlation between
the fraction of dark matter within $R_e$ and the logarithmic value of
$R_e$. With our measured value of $R_e$, this correlation predicts 
a dark matter fraction at $R_e$ of $0.68$.  Our value of $0.52$ is smaller,
but still within their 68\% confidence limit.

\citet{tho09} derive scaling relations for halo parameters based on 
observations of
early-type galaxies in the Coma cluster.  These relations are constructed 
for similar galaxies using the same halo parameterization and
modeling code used in this paper.  This makes comparison to our parameters 
straightforward.  
We compare to the observed relations between halo parameters $r_c$, $V_c$, 
and $\rho_c$ and total blue 
luminosity $L_B$.  Our value of $V_c$ falls directly on the $V_c$-$L_B$ 
relation, however our measured $r_c$ is smaller by roughly an order of 
magnitude.  Since $\rho_c \propto V^2_c / r^2_c$, the discrepancy in 
$r_c$ causes our measurement of $\rho_c$ to be high when compared to 
the \citet{tho09} $\rho_c$-$L_B$ relation.  Scatter in this relation is large,
however, and the environment of NGC~4594 is different from that of the Coma
galaxies.

\citet{kor04b, kor11b} also derive scaling laws for similar parameters in
galaxies of later Hubble type (Sc-Im).  We measure a much higher density
and much smaller core radius than the \citet{kor04b,kor11b} 
relations imply at the $L_B$ of NGC~4594.  We interpret this as the result of
severe compression of the halo by the gravity of the baryons \citep{blu86}.
Such an effect is expected in early-type galaxies with massive bulges.

\begin{acknowledgements}

We thank Eric Emsellem for providing reduced SAURON data and helpful comments.
This work would not be feasible without the excellent resources of the
Texas Advanced Computing Center (TACC). KG acknowledges support from NSF-0908639.  DR is grateful for hospitality and support from the Institute for Advanced
Study in the form of a Corning Glass Works Foundation Fellowship.

\end{acknowledgements}

\bibliographystyle{apj}

\end{document}